\documentstyle[prd,preprint,aps,floats,epsfig]{revtex}

\begin{document}

\newcommand{\be}{\begin{equation}}
\newcommand{\ee}{\end{equation}}
\newcommand{\bea}{\begin{eqnarray}}
\newcommand{\beas}{\begin{eqnarray*}}
\newcommand{\eea}{\end{eqnarray}}
\newcommand{\eeas}{\end{eqnarray*}}
\newcommand{\ba}{\begin{array}}
\newcommand{\ea}{\end{array}}
\def\ls{\mathrel{\lower4pt\vbox{\lineskip=0pt\baselineskip=0pt
           \hbox{$<$}\hbox{$\sim$}}}}
\def\gs{\mathrel{\lower4pt\vbox{\lineskip=0pt\baselineskip=0pt
           \hbox{$>$}\hbox{$\sim$}}}}

\def\nt#1{\not{\hspace*{-.7ex}{#1}\hspace*{.5ex}}}


\tightenlines

\def\DESepsf(#1 width #2){\epsfxsize=#2 \epsfbox{#1}}
\draft
\thispagestyle{empty}
\preprint{\vbox{ \hbox{UMD-PP-03-029}
\hbox{December, 2002}}}

\title{Neutrino Mass, Proton Decay and Dark Matter
in TeV Scale Universal Extra Dimension Models}

\author{\large
R.N. Mohapatra$^1$\footnote{e-mail:rmohapat@physics.umd.edu} and
A. P\'erez-Lorenzana$^{2,3}$\footnote{aplorenz@fis.cinvestav.mx}}

\address{$^1$Department of Physics, University of Maryland\\
College Park, MD 20742, USA\\
$^2$The Abdus Salam International Centre for Theoretical Physics,
I-34100, Trieste, Italy\\
$^3$Departamento de F\'\i sica,
Centro de Investigaci\'on y de Estudios Avanzados del I.P.N.\\
Apdo. Post. 14-740, 07000, M\'exico, D.F., M\'exico}

\maketitle

\thispagestyle{empty}

\begin{abstract}
We show how the problem of small neutrino masses and suppressed
proton decay can be simultaneously resolved in 6-D universal extra
dimension models (UED) with a low fundamental scale using extended
gauge groups that contain the local $B-L$ symmetry. The extra
space dimensions are compactified either on a $T^2/Z_2$ or
$T^2/Z_2\times Z'_2$ orbifold depending on whether the full gauge
group is $SU(2)_L\times U(1)_{I_{3R}}\times U(1)_{B-L}$ or
$SU(2)_L\times SU(2)_{R}\times U(1)_{B-L}$. In both cases,
neutrino masses are suppressed by an appropriate orbifold parity
assignment for the standard model singlet neutrinos and the proton
decay rate is suppressed due to a residual discrete symmetry left
over from compactification. For lower values of the fundamental
scale, a dominant decay mode of the neutron is $n\rightarrow 3
\nu$. An interesting consequence of the model is a possible two
component picture for dark matter of the universe.
 \end{abstract}

\section{Introduction}

The possibility that the fundamental scale of nature ($M_*$) is in
the multi-TeV range has been the focus of great deal of activity in
the past several years~\cite{arkani}. These models are inspired by
results in nonperturbative string theories and may therefore be
providing a possibility of testing string theories in colliders as
well as other low energy measurements. Furthermore, they also
provide a new way to look at the puzzles of the standard model such
as the gauge hierarchy and fermion masses. Even regardless of these
motivations, their sharp differences from the usual ``grand''
desert picture of beyond the standard model physics makes them
interesting enough to pursue their phenomenological as well as
cosmological implications. As it is well known the high value of
the conventional Planck scale is a derived scale in these models
and arises out of a combination of $M_*$ and the compactification
sizes of the extra dimensions. When the sizes of the extra
dimensions are large, many other phenomenological consequences
ensue and have been widely discussed in literature.

Even though these models are attractive in many ways, they have two
fundamental problems that need to be resolved before they can be taken
seriously: (i) they provide no simple way to understand
small neutrino masses since the familiar seesaw mechanism\cite{seesaw}
requires scales of order $10^{12}$ GeV or so, which are much higher than
$M_*$; and (ii) no simple way to suppress proton decay induced by
physics at the string scale.

A proposal for solving the neutrino mass problem using a singlet
neutrino in the entire 5-dimensional bulk (so called bulk
neutrinos) was proposed in ref.~\cite{dienes}. The smallness of
neutrino masses in this picture owe their origin to large bulk
radius rather than a large mass scale as in seesaw models. This
new way of
 discussing neutrino masses has many new implications which have been
studied in several papers~\cite{3nus}. In this case however, one
has to assume the existence of a global B-L symmetry  in the
theory to prevent dangerous operators like $(LH)^2/M_*$ from
destabilizing the neutrino masses. Since there is a general lore
that there should be no global symmetries in string theories,
barring the possibility that the global symmetry arises
accidentally, we face a problem.

Proton decay in such models can arise due to the presence of
nonrenormalizable operators of the form
 \be
 \label{dim6op}
 { Q Q Q L \over M_*}~;
 \ee
which would clearly lead to an unacceptably short lifetime for the proton,
many orders of magnitude below the present experimental lower limit.

In a recent paper~\cite{ponton} a solution to the proton decay
problem was proposed in the context of the so called universal
extra dimension models (UED)~\cite{cheng} where the number of
space-time dimensions where all standard model (SM) fields reside is six
and the fundamental scale of nature is in the TeV range. The main
observation of~\cite{ponton} is that in six
dimensional UED models, the extra space-time dimensions (the 5th
and 6th dimensions) provide a new $U(1)$ symmetry under which the
SM fermions are ``charged'' and enough of this symmetry survives
the process of orbifold compactification that it suppresses proton
decay to a very high degree. In the model of \cite{ponton}, the
surviving symmetry is $Z_8$, so that the leading order baryon
number violating operator has dimension 16 in six dimensions and
thus highly suppressed.

No simple solution to the neutrino mass problem has been found in these
models.

In a recent paper\cite{5dlrm}, a new way to solve the neutrino mass
problem in TeV scale gravity models was proposed in higher
dimensional theories without invoking the bulk neutrino
or large extra dimensions. Instead, it was noted that in a
5-dimensional UED model based on an extended gauge symmetry
$SU(2)_L\times SU(2)_R \times U(1)_{B-L}$ if the right handed
neutrino is given a ``twisted'' orbifold boundary condition, the
leading order neutrino mass operator has sufficiently high
dimension so that its contribution to $m_\nu$ is suppressed
despite the fundamental scale being low. The essential
trick used is to project the familiar
 right handed neutrinos out of the zero mode spectrum of the theory.
The main reasons that this solves the neutrino mass problem
are as follows:

\begin{itemize}

\item (i) the local B-L symmetry forbids the operator $LHLH/M^*$;

\item (ii) the assignment of the orbifold symmetry to the
chiral components of the right handed neutrino $N$ forbids the
conventional Dirac mass of the neutrinos.

\end{itemize}

The resulting lowest dimension operator contributing to the Majorana
neutrino mass in five dimensions has dimension d=10 and there is a
mini-seesaw for the Dirac mass leading to eV neutrino mass without any
fine tuning. There was also no
need to assume any global symmetries. A fundamental scale $M_*$ of 30-100
TeV and a compactification scale of order of a TeV was sufficient for
phenomenological consistency as well as for giving small neutrino
masses. It was speculated in this paper that when
extended to six dimensions, the model also solves the proton decay
problem.

 In this paper, we first present a simpler six dimensional model based on
the smaller gauge group $SU(2)_L\times U(1)_{I_{3R}}\times
U(1)_{B-L}$ and a more economical fermion spectrum and show using
$T^2/Z_2$ compactification that, we can solve both the neutrino
mass and proton decay problems simultaneously. The
compactification scale as well as the scale of $B-L$ breaking in
this model are in the range of a TeV and the fundamental scale
$M_*$ anywhere from 10-100 TeV. We then extend the  gauge group to
the left-right symmetric case and show how the same results are
maintained; in this case however, we need to have the
compactification on $T^2/Z_2\times Z'_2$, which is very similar to
our earlier work~\cite{5dlrm}, now extended to six dimensions.
Furthermore, we find some new decay modes for the nucleon in both
models that may have lifetimes in the range of experimental
accessibility.  We  then discuss the question of dark matter in
these models.

In addition to solving the neutrino mass and proton decay problem, the
models have the following interesting features:

\begin{itemize}

\item they predict the existence of a light $Z'$ with mass at the
compactification scale (of order of a TeV);

\item they predict baryon number violating processes where both
neutron and proton  decay to final states with three leptons, e.g.
 $n\rightarrow \nu \bar\nu_s \bar\nu_s$;~
$n\rightarrow \pi^0 \nu_e \bar\nu_s \bar\nu_s$;~
$n\rightarrow \pi^+  e^- \bar\nu_s \bar\nu_s$;~ and
$p\rightarrow \pi^+ \nu_e \bar\nu_s \bar\nu_s$.

\item The model gives a two component picture of dark matter with
$\gamma_{KK}$ and $\nu_R$ both playing the role of dark matter.

\end{itemize}

We organize our discussion as follows: in section II we present
the $SU(2)_L\times U(1)_{I_{3R}}\times U(1)_{B-L}$ model and show
how one solves the neutrino mass and proton decay problems using
$T^2/Z_2$ compactification. In sec. III, we introduce the
left-right extension of this model.
 In section IV, we discuss the anomaly
cancellations conditions for both these models, which drive us to
the conclusion that the model should contain at least three
families. This provides a generalization of the results of
Ref.~\cite{poppitz} to the case of extended gauge groups. In
section V, we discuss the $T^2/Z_2\times Z'_2$ orbifolding of the
two extra space-like dimensions, the breaking of the six
dimensional Lorentz symmetries and the generalities for the mode
expansion and mass spectrum  of the theory. In section VI we
introduce our $Z_2\times Z'_2$ charge assignments for the
particles of the model and discuss the orbifold and spontaneous
symmetry breaking. The mass pectrum and a brief discussion of
phenomenology is the subject of sec. VII. We use the previous results to
analyze
the origin of the neutrino mass for the left-right model in section
VIII. In section IX, we systematically analyze the baryon violating
operators to study proton and neutron decay on the model. In sec. X,
we discuss the two component picture of the dark matter in the universe.
We present our conclusions in section XI. Some useful results are
finally given in the appendix.

\section{$SU(2)_L\times U(1)_{I_{3R}}\times U(1)_{B-L}$ model in 6-D,
neutrino mass and nucleon decay}

In this section, we consider a six dimensional model based on the
gauge group $SU(2)_L\times U(1)_{I_{3R}}\times U(1)_{B-L}$, with
the six dimensional gravitational anomaly free particle content as
in \cite{ponton}: $Q_{-}(2,0,1/3),~ \psi_{-}(2,0,-1),~
 U_{+}(1,1/2,1/3),~D_{+}(1,-1/2,1/3),~E_{+}(1,-1/2,-1)$,
 $~N_{+}(1,+1/2,-1)$ where $Q^T=(u,d)$ and $\psi^T=(\nu, e)$ and the
subscripts $\pm$ denote the six dimensional chirality; the numbers
in the parentheses are the gauge quantum numbers.
The
corresponding six dimensional chirality projection operator is
defined as $P_\pm = (1 \pm \Gamma^7)/ 2$, where $\Gamma^7$ is
itself given by the product of the six (eight by eight) Dirac
matrices: $\Gamma^7= \Gamma^0\Gamma^1\cdots\Gamma^5$.
Note that
the right handed neutrino is required for cancellation of
gravitational anomalies in six dimensions. We will denote the
space coordinates by $(x^0,x^1, x^2, x^3, x^4, x^5)$ and often
write $x^4=y_1$ and $x^5=y_2$.

Each fermion field in the above equation is a four component field
with two 4-dimensional 2 component spinors with opposite chirality
e.g. $Q_-$ has a left chiral $Q_{-,L}$ and a right chiral field
$Q_{-,R}$ (see appendix). As such the theory is vectorlike at this
stage and we will need orbifold projections to obtain a chiral
theory.

We compactify the theory on a $T^2/Z_2$ orbifold; where $T^2$ is
defined by the extra coordinates $y_{1,2}$ satisfying the following
conditions: $y_{1,2} = ~y_{1,2}+2\pi R$ and $Z_2$ operates on the
two extra coordinates as follows: $(y_1,y_2)\rightarrow
(-y_1,-y_2)$. We now impose the orbifold conditions on the fields
as follows: we choose the following fields to be even under the
$Z_2$ symmetry: $Q_L,\psi_L, U_R, D_R, E_R, N_L$; the kinetic
energy terms then force the opposite chirality states to be odd
under $Z_2$. Note specifically that, along with the $SU(2)_L$
singlet fields $U_R,D_R,E_R$, the $N_L$ is chosen even under $Z_2$
instead of the $N_R$ field. This is crucial to our understanding of
neutrino masses. In usual extensions of the standard model to
incorporate neutrino masses, one usually includes the $N_R$ fields.
If one instead included the $N_L$ field, theory would have been
anomalous. In our case however, since the zero modes descend from
an anomaly free higher dimensional theory, this problem does not
arise. In fact, the apparent anomaly in the zero mode sector of the
theory would be cancelled  by appropriate Chern-Simon terms that
would be induced in the process of compactification and by the
Green-Schwarz mechanism~\cite{gs}.

 As is well known, the even fields when Fourier expanded
involve only the $cos \frac{\vec{n}\cdot\vec{y}}{R}$ and the odd
fields only $sin \frac{\vec{n}\cdot\vec{y}}{R}$; for $\vec{n}=(n_1,
n_2)$ a pair of integer numbers. As a result only the $Z_2$ even
fields  have zero modes. Thus, with the above compactification,
below the mass scale $R^{-1}$, the only fermionic modes are those
of the standard model plus the sterile neutrino
 $\nu_s\equiv N^{0}_L$.

The gauge group below this energy is the entire gauge group of the
theory i.e. $SU(2)_L\times U(1)_{I_{3R}}\times U(1)_{B-L}$. To
implement gauge symmetry breaking, we choose one Higgs doublet
$\phi(2,-1/2,0)$ with $B-L=0$ and a singlet $B-L =-1 $ Higgs boson
$\chi(1,1/2,-1)$. We choose both the Higgs fields to be even under
the $Z_2$ symmetry. We will use their zero modes to break the
gauge symmetry as in the standard model. When we give vev to the
field $\langle\chi\rangle =~v_{BL}$, it breaks the group down to
the standard model. We will choose $v_{B-L}\sim 800$ GeV to a TeV.

Before discussing the implications of the model for neutrino masses and
proton decay, let us study the extra symmetries of the 4-dimensional
theory implied by the fact that it derives from a 6-dimensional one.

First, a discrete translational symmetry insures the conservation
of the fifth and sixth momentum components, $p_a$, which are
quantized in integer factors of $1/R$.

Secondly, in the full uncompactified six dimensional theory, there
is an extra $U(1)_{45}$ symmetry associated with the rotations in
the $x_4$-$x_5$ plane. After compactification, the $U(1)_{45}$
invariance reduces to a $Z_4$ symmetry. Therefore invariance under
the $SO(1,3)\times Z_4$ space-time Lorentz transformations must be
imposed on all possible operators allowed in the effective four
dimensional theory, i.e. the allowed operators will be those that
are invariant under the whole $SO(1,5)$ symmetry, plus those for
which the sum of fermion $U(1)_{45}$ charges equals zero modulus 8.
The reasoning is as follows: the $Z_4$  spatial symmetry, actually
translates into a $Z_8$ symmetry group for the spinorial
representation. In fact under a $\pi/2$ rotation of the $x_4$-$x_5$
plane a fermion transforms as $\Psi(x') = U\Psi(x)$; with $U=
\exp[i(\pi/2)\Sigma_{45}/2]$; where  $\Sigma_{45}=
i[\Gamma^4,\Gamma^5]/2$ is the generator of the $U(1)_{45}$ group
(see appendix for details).

To see which operators are allowed, we need to know the $U(1)_{45}$
quantum numbers of the theory which can be easily read of from the
six dimensional theory and are given in  Table I.

\begin{table}
\begin{tabular}{|c||c||c|}\hline
Fields & $Z_2$ parity & $U(1)_{45}$ charge \\ \hline
$Q_L,\psi_L$ &+ & +1/2\\\hline
$U_R,D_R,E_R $& + & +1/2 \\ \hline
$N_L$ & + & -1/2 \\ \hline
$Q_R,\psi_R $ & -& -1/2 \\ \hline
$U_L, D_L, E_L$ & - & -1/2 \\ \hline
$N_R$ & - & +1/2 \\ \hline
\end{tabular}
\caption{$U(1)_{45}$ and orbifold $Z_2$ assignment quantum numbers of the
fermion fields in the model}
\end{table}

It may be helpful to note that the rule for $U(1)_{45}$ charges of
various fermion fields is that for $\Psi_{\pm,L}$, it is
$\mp\frac{1}{2}$ and for $\Psi_{\pm,R}$, it is $\pm\frac{1}{2}$
(see appendix). We will use these quantum numbers below in
constructing all allowed higher dimensional operators, which must
conserve the $U(1)_{45}$ charge modulus 8.

\subsection{Neutrino mass}
Note that in this model due to our orbifold assignments and choice
of the gauge group coupled with the residual $Z_8$ symmetry
discussed above, we only have one term that contributes to neutrino
masses in the leading order. The leading order allowed term is:
 \be
 \lambda \frac{\psi^T_L C^{-1}N_L\phi
 (\chi^*)^2}{M^5_*}.
 \label{nuop}
 \ee

The following potentially dangerous terms are forbidden by the symmetries
of the 6-dimensional theory:

\begin{itemize}

\item $(\psi_L\phi)^2/M_*$  is forbidden by $B-L$ symmetry.

\item Terms like $\bar{\psi}_L\phi N_R$,
though allowed are not problematic
due to $Z_2$ quantum numbers which imply that the $N_R$ has no zero mode.

\item $\frac{(\psi_L\phi)^2(\chi^*)^2}{M^7_*}$ and $N_LN_L(\chi^*)^2$ are
forbidden by the residual $Z_8$ symmetry.

\end{itemize}

The operator allowed by the symmetries are written in the
6-dimensional field theory. Upon compactification to the
4-dimensions, it reduces to the form $\lambda \frac{\psi^{(0)T}_L
C^{-1}N^{(0)}_L\phi^{(0)} (\chi^{(0)*})^2}{M^5_*R^3}$ (where the
superscript $0$ denotes the zero mode of the field). Using
$M_*\simeq 100$ TeV and $R^{-1}\sim $ TeV and using $\lambda \sim
0.1$, we find for that it leads to $m_\nu\sim $ eV, which is in the
right range to fit oscillation data without any fine tuning. If we
allowed $\lambda \approx 10^{-6}$, the value of the scale $M_*$
could be 10 TeV. Furthermore, the neutrinos in this model are Dirac
particles since all Majorana terms are forbidden by the $Z_8$
symmetry. This is basically due to the fact that charge conjugation
and chirality operators now commute (see also the appendix).
 As a result, neutrinoless double beta decay is forbidden in this
model.

\subsection{Baryon nonconservation}
Let us now study the consequences of the extra local $B-L$ symmetry
and the 6-dimensional geometry for baryon nonconservation. First,
it is easy to check that the operator in Eq.~(\ref{dim6op}) is not
invariant under the residual $Z_8$ symmetry of the orbifold.
Indeed, considering again that all the SM fields are zero modes
that have a $Z_8$ charge $+1$, one concludes that such an operator
has $\Delta\Sigma_{45} = 4$ (notice that only the zero mode fields
are relevant for B-violating processes of experimental interest).
Furthermore, an operator with $\Delta B\neq 0$ has at the zero mode
level \be \Delta \Sigma_{45} = 3~\Delta B + \Delta L_{SM} - \Delta
L_{\nu_s} \label{ds} \ee where $\Delta L_{SM}$ gives the total
lepton number in SM fields whereas $\Delta L_{\nu_s}$ gives the
sterile ($\nu_s \equiv N_L$) contribution to  lepton number. In
terms of the $B-L$ gauge charge, one can write above equation in
the form: $\Delta \Sigma_{45} = 3 \Delta (B-L)_{quarks} - \Delta
(B-L)_{leptons} + \Delta (B-L)_{\nu_s}$.

Due to the Lorentz invariance condition in Eq.~(\ref{ds}), the
simplest $\Delta B=1$ operators must involve at least three quarks
and three leptons. One can further classify all non renormalizable
six fermion $\Delta B=1$ couplings according to their sterile
neutrino content. One then has operators with $\Delta L = 3$ for
those involving three $\nu_s$'s; $\Delta L = 1$ for  two; $\Delta L
= -1$ for  one; and finally and $\Delta L = -3$ for operators non
involving any  $\nu_s$ at all. Clearly, only those with $\Delta L =
1$ are naturally invariant under $B-L$. All others need to involve
at least two $\chi$ scalar fields to compensate the $B-L$ charge.
Hence, any of the lowest dimension operators  should have two
sterile neutrinos at the zero mode level. Here are some typical
operators involving the zero mode fields which arise after orbifold
compactification from the allowed operators in six dimensions:
 \bea
 a) &\qquad & (\bar{\psi}_L d_R)( Q^T_LC^{-1} N_L) ( d^T_R C^{-1} \nt D
N_L)~; \nonumber \\
 b) && (\bar \psi_L\nt D Q_L)
     ( d^T_RC^{-1} \gamma^\mu N_L)( d^T_R C^{-1}\gamma_\mu
N_L)~; \nonumber \\
 c) && (\bar e_R \nt D d_R)
   ( d^T_RC^{-1} \gamma^\mu N_L)(d^T_RC^{-1}\gamma_\mu N_L)~;
 \label{dim16eff}
 \eea
respectively, with an overall strength of
 $\sim [M_*^6 (M^*\pi R)^4]^{-1}$.
It is worth noticing that these operators are completely different
from those considered in Ref.~\cite{ponton} for the 6D standard
model.

Operators in Eq.~(\ref{dim16eff}) contribute to the following
processes:
An invisible neutron decay: $n\rightarrow\nu_L\bar\nu_s\bar\nu_s$.
Nucleon, four body decays: $n\rightarrow\pi^0\nu_L\bar\nu_s\bar\nu_s$ and
$p\rightarrow\pi^+\nu_L\bar\nu_s\bar\nu_s$; and  five body decays:
$n\rightarrow\pi^0\pi^0\nu_L\bar\nu_s\bar\nu_s$,
$n\rightarrow\pi^+\pi^-\nu_L\bar\nu_s\bar\nu_s$ and
$p\rightarrow\pi^+\pi^0\nu_L\bar\nu_s\bar\nu_s$.
They also contribute to
$n\rightarrow\pi^+\ell^-\bar\nu_s\bar\nu_s$;
$n\rightarrow\pi^+\pi^0\ell^-\bar\nu_s\bar\nu_s$ and
$p\rightarrow\pi^+\pi^+\ell^-\bar\nu_s\bar\nu_s$.
The overall amplitude for these processes,
(up to possible coupling constants and form and phase space factors)
gives a decay rate  of the order
 \be
 \Gamma_{dim16} \sim
 {1\over {3}^{13} (\pi R M_*)^8}\left( {m_N\over M_*}\right)^{12} m_N~;
 \ee
with $m_N$ the mass of the nucleon used here to fix the scale (as
$m_N/3$. The powers of $3^{13}$ are due to the fact that the
nucleon undergoes a three body decay.

It turns out that in this model there are next order operators, which in
some cases are less suppressed. Examples of such
 {\it dimension 17} operators are (in terms of the zero mode fields)
 \bea
a) &\qquad& (\bar\psi_L  \gamma^{\mu}  Q_L)
 (d^T_R C^{-1}\gamma_{\mu} N_L)
 ( Q_L^T C^{-1} N_L)\phi~; \nonumber \\
b)  &&  (Q^T_L C^{-1}N_L)^2(\bar\psi_L d_R)~ \phi~;
 \label{dim17}
 \eea
These operators induce the very same processes
as those already mentioned above. The decay rates in this case are of
order
 \be
 \Gamma_{dim17} \sim
 {v_{wk}^2\over  3^{11}M_*^2(\pi R M_*)^{10}}
 \left( {m_N\over M_*}\right)^{10} m_N \sim
 {9 v_{wk}^2 \over (\pi RM_*)^2 m_N^2}~ \Gamma_{dim16}~.
 \ee
The overall factor on the right hand side of this equation
appears due to the replacement of the  $(m_N/3)^2$ contribution
from the covariant derivative in Eq.~(\ref{dim16eff}) by
$v_{wk}^2/(\pi R M_*)^2$. For small gaps between  compactification
and fundamental scales ($\pi R M_*<100$),
the  last factor gets larger, and
thus the contribution of the dimension 17 operators becomes the leading
order.
A simple estimation  gives the lifetime
\be
\tau \approx ~6\cdot 10^{30}~{\rm yr}\times
\left[{10^{-4}\over\Phi_n}\right]
 \left({\pi RM_*\over 10}\right)^{10}~
 \left({M_*\over 10~{\rm TeV}}\right)^{12}~;
\label{tau}
\ee
where we have explicitly introduced the  contribution of the kinematical
phase space factor, $\Phi_n$, which depends in the specific process with
$n$
final states. (A possible order one form factor which enters in
the case of two pion production has not been written.)

The simplest process we have is
$n\rightarrow  \nu_L\bar\nu_s\bar\nu_s$, which has three final states,
and so $\Phi_3\ls (4\pi)^{-3} ~O(10^{-2})$.
Hence, the theory is quite safe in this regard.
Actually, with the values we used for getting a `natural'
neutrino mass one gets a very large bound for
all nucleon  decay modes: $\tau\gs 10^{48}$~yr.
Nevertheless, if one allows a soft hierarchy, say having $\lambda \sim
10^{-6}$,
one can take the values suggested in Eq.~(\ref{tau}),
 $M_*\sim 10~{\rm TeV}$ and
$\pi RM_*\sim 10$, thus getting lifetimes just on the edge of
present experimental limits. For comparison, search for the decays
$p\rightarrow e^-\pi^+\pi^-$; and $n\rightarrow e^-\pi^+$ set
limits in about $\tau_p>3\cdot10^{31}$~yr~\cite{frejus} and
$\tau_n>6.5\cdot10^{31}$~\cite{seidel} respectively. There is a
proposal to search for the decay mode $n\rightarrow 3 \nu$ in the
KAMLAND experiment\cite{kamyshkov}.

\section{Left-Right Symmetry in 6D}

In this section, we extend the discussion of the previous section to the
 left-right symmetric model based on the gauge group $SU(2)_L\times
SU(2)_R\times U(1)_{B-L}$, which contains the subgroup  $SU(2)_L\times
U(1)_{I_{3R}}\times U(1)_{B-L}$. In addition to the aesthetically
appealing feature
of having parity as an asymptotic symmetry in this theory, it provides new
phenomenology associated with the TeV scale right handed $W^\pm_R$, which
may be accessible to colliders. A 5-dimensional left-right model was
discussed in \cite{5dlrm}. The six dimensional left-right model shares
many of the features of the 5-dimensional model- e.g. the smallness of
neutrino mass, although the details are different; low scale $W_R$ and
$Z'$ boson and an effective model below the compactification scale based
on  $SU(2)_L\times U(1)_{I_{3R}}\times U(1)_{B-L}$ with a light sterile
neutrino. There are however several new aspects that we discuss now.

The major new points are three fold: (i) in contrast with the
$SU(2)_L\times U(1)_{I_{3R}}\times U(1)_{B-L}$ model, in this case
the standard model fermion spectrum requires that the orbifold
compactification be made on a $T^2/(Z_2\times Z'_2)$ space; (ii) in
contrast with the 5-dimensional left-right model, now the proton
decay problem can be solved using the $U(1)_{45}$ symmetry as in
the previous section. This provides a simultaneous resolution of
both the neutrino mass and proton decay problem with a TeV scale
gravity. Since the compactification in our case is different
 from that in \cite{ponton}, in order to show we indeed solve the proton
decay problem, one has to show that the
residual $Z_8$ symmetry survives at low energies; (iii) a discussion of KK
dark matter particle, where we show that in this model dark matter has two
cold components: the $\gamma^1_{KK}$ and $N_R$. For the sake of
completeness, we review some of the salient points of the model given in
Ref.\cite{5dlrm} when extended to the 6-D case.

To discuss the model further, we denote the gauge bosons as $G_M$;
$W^{\pm,0}_{1,M}$; $W^{\pm,0}_{1,M}$; and $B_M$, for $SU(3)_c$,
$SU(2)_L$, $SU(2)_R$ and $U(1)_{B-L}$ respectively, where
$M=0,1,2,3,4,5$ denotes the six space-time indices. We will also
use the following short hand notations:  Greek letters
$\mu,\nu,\dots=0,1,2,3$ to denote  usual four dimensions indices,
as usual, and lower case Latin letters $a,b,\dots=4,5$ for those of
the extra dimensions. We will also use $\vec y$ to denote the
($x_4,x_5$) coordinates of a point in the extra space.

For matter content, we choose
four  quark  and four lepton
representations per generation as follows:
 \bea
 {\cal Q}_{1,-}, ~{\cal Q}'_{1,-}= (3,2,1,1/3)~; &&
 {\cal Q}_{2,+}, ~{\cal Q}'_{2,+}= (3,1,2,1/3)~;\nonumber\\
 \!\!{\cal \psi}_{1,-}, ~{\cal \psi}'_{1,-}= (1,2,1,-1)~; &\quad \quad&
 ~{\cal \psi}_{2,+}, ~{\cal \psi}'_{2,+}= (1,1,2,-1)~;
 \label{matter}
 \eea
where,  within brackets, we have written the quantum numbers that
correspond to each group factor, respectively. Notice that we have
duplicated the spectrum with respect to the usual four dimensional
left right model. One needs to do so in order to reproduce the
standard model (SM) content in the four dimensional theory, as it
will become clear later on when discussing the orbifolding on the
theory. We will also assume all above matter fields to be chiral in
the six dimensional sense. Notice  we have chosen all fields with
subscript `1' (`2') to have a positive (negative)  six dimensional
chirality. Thus,  the matter content of the theory would be
symmetric under the subscript interchange: $1\leftrightarrow 2$.

\section{Cancellation of anomalies and the number of generations}

With the above assignments the model describes chiral interactions
that should be made anomaly free to be consistent.
There are two classes of  anomalies:
{\it local} and {\it global} anomalies
(for a discussion see~\cite{Gouverneur}).
Local anomalies are related to
infinitesimal  gauge and/or  coordinate transformations,
whereas global anomalies are essentially nonperturbative.

In six dimensions, local anomalies arise from box one-loop diagrams
where the external legs are either gauge bosons or gravitons.
Diagrams with only gauge bosons in the external legs
correspond to the pure gauge anomaly, whereas those with only gravitons
give the pure gravitational anomaly.
Diagrams with both gauge bosons and gravitons
correspond to mixed anomalies.

In our model $U(1)_{B-L}$ is vector-like due to the replication of
representations with opposite chiralities. Same happens for color
$SU(3)_c$ group. Thus, $SU(3)_c\times U(1)_{B-L}$ anomalies cancel
within each generation. Same holds for the subgroup $U(1)_Q$. In
fact, the model has no irreducible local gauge anomalies. The only
possible anomalies of this kind, which are $[U(1)]^4$ and
$[SU(3)]^3 U(1)$ vanish identically. All other anomalies associated
to: $[SU(2)_{1,2}]^4$; $[SU(3)]^2[SU(2)_{1,2}]^2$; and
$[SU(2)_{1,2}]^2 [U(1)]^2$; are reducible. They are not a matter of
concern, because they  can be cancelled through the Green-Schwarz
mechanism~\cite{gs} by the introduction of an appropriate set of
two index antisymmetric tensors. The presence of reducible
anomalies is rather generic in six dimensional chiral theories,
thus, antisymmetric tensor are likely to be an ingredient of any
six dimensional model (see for instance the  models in
Refs.~\cite{cheng,Gouverneur,piai}). Notice that all local
anomalies that are cubic in $SU(2)_{1,2}$ are identically zero.

Regarding our first model, we should note that all the above
arguments also hold since $U(1)_{I_3}$ is actually the diagonal
subgroup of the $SU(2)_2$ group, whereas the matter content is
identical (up to a replication). Same would be true for the rest of
the discussion alone this section.

As the total number of fermions with chirality $+$ is equal to the
number of fermions with chirality $-$, there is no pure
gravitational anomaly. Regarding mixed anomalies, only those
associated to  diagrams with two gravitons in the external legs can
be non zero~\cite{gaume}. Again, due to the vector-like nature of
$U(1)_{B-L}$ and $SU(3)$ such anomalies vanish for these groups.
Mixed anomalies that involve $SU(2)_{1,2}$ are all reducible, and
cancelled by the same tensors that take care of the reducible pure
gauge anomalies.

Global anomalies are, on the other hand,
more restrictive for the fermion content of the model.
These anomalies are related to local symmetries that can not be deduced
continuously from the identity.
Cancellation of the of global gravitational anomalies  in six dimensions,
however, is
automatically insured by the cancellation of the local gravitational anomaly.
Therefore, only global gauge anomalies are  possible.
In general, they are associated to a non trivial topology of the gauge group.
Particularly, they arise in six dimensional theories
when the sixth homotopy group, $\pi_6$,
of the gauge group $G$ is non trivial ($\pi_6(G)\neq 0$).
Cancellation of such an anomaly needs an appropriate matter content.
As a matter of fact, they may occur for $SU(3)$
as well as $SU(2)$ gauge theories~\cite{witten,vafa}.
Given that $\pi_6[SU(3)]= Z_{6}$  and $\pi_6[SU(2)]= Z_{12}$,
the  cancellation of the  global gauge anomalies constrains the number of
chiral triplet color  representations
in the model, $N_c(3_\pm)$, to satisfy:
 \be
 N_c(3_+) - N_c(3_-) = 0 ~~\mbox{mod}~6 ~
 \ee
As $SU(3)$ is vector like  this condition is naturally fulfilled.
For the number of $SU(2)_{1,2}$ chiral doublets, $N_{1,2}(2_\pm)$,
it also  requires that
 \be
 \label{cons}
 N_{1,2}(2_+) - N_{1,2}(2_-) = 0 ~~\mbox{mod}~6 ~.
 \ee
Last condition indicates that the global anomaly does not cancel
within a single family, because for both the $SU(2)$ their eight
fermion representations are all of the same chirality, either $-$
for $SU(2)_1$ or $+$ for $SU(2)_2$. Looking at the matter content
in Eq.~(\ref{matter}) one easily sees that the above constraint can
be written in a unique way in terms of the number of generations,
$n_g$, which is the number of exact replications of our matter
content,  as follows:
 \be
 n_g =0~~\mbox{mod}~ 3~.
  \ee
Hence, 3 is the minimal number of generations for which the theory
is mathematically consistent. This is a remarkable result. It  was
already known for the case of the six dimensional extension of the
Standard Model~\cite{poppitz}, and it nicely remains in the present
left-right extension. This is also true for the case of the
$SU(2)_L\times U(1)_{I_{3R}}\times U(1)_{B-L}$ model where the same
condition (\ref{cons}) holds.

\section{Orbifolding}

\subsection{ $T^2/Z_2\times Z'_2$ orbifolding,
and its space-time symmetries}

As already discussed, in six dimensions, the chiral spinors are vector
like in four dimensions. In order to get a chiral theory, one must do
appropriate projection. We discuss this below.

{}First, we compactify the extra $x_4$, $x_5$ dimensions
into a torus, $T^2$, with equal
radii, $R$, by imposing  periodicity conditions,
$\varphi(x_4,x_5) = \varphi(x_4+ 2\pi R,x_5) = \varphi(x_4,x_5+ 2\pi R)$
on any field $\varphi$.
The physical space for this manifold is then represented by the squared
interval: $[-\pi R,\pi R]\times[-\pi R,\pi R]$.
This has the effect of breaking the original $SO(1,5)$
Lorentz symmetry group of the six dimensional space  into
the subgroup $SO(1,3)\times Z_4$, where the last factor corresponds to the
group of discrete  rotations in the $x_4$-$x_5$ plane,
by angles  of $k\pi/2$ for $k=0,1,2,3$.
This is a subgroup
of the continuous $U(1)_{45}$ rotational symmetry contained in $SO(1,5)$.
The remaining $SO(1,3)$ symmetry  gives the usual 4D Lorentz invariance.
Notice also that due to the periodicity conditions, the Poincare translational
invariance remains unbroken. In fact, the center of the square
could have been
chosen anywhere on the torus, thus  given the same physical space
up to a  coordinate redefinition.

Next, we double orbifold the torus by requiring that the theory be invariant
under the transformations:
 \be
 \label{z}
 Z_2: \vec{y} \rightarrow -\vec {y} \qquad \mbox{and}\qquad
 Z'_2: \vec{y}~' \rightarrow -\vec{y}~'
 \ee
for $\vec y = (x_4,x_5)$; and
where $\vec{y}~' = \vec{y} - (\pi R /2, \pi R/2)$.
As it is shown in figure 1, this orbifolding
has four fundamental fixed points that bound the
fundamental space, which is now  reduced to a smaller
squared interval that we identify with: $[0,\pi R]\times[-\pi R/2,\pi R/2]$.
The fundamental $Z_2$ fixed points, $O_{1,2}$,
are then located at the coordinates
$(0,0)$ and  $(0,\pi R)$, whereas those of $Z'_2$,
$O_{3,4}$, are located at the points
$(\pi R/2,\pm \pi R/2)$, respectively.

\begin{figure}[ht]
\centerline{
\epsfysize=150pt
\epsfbox{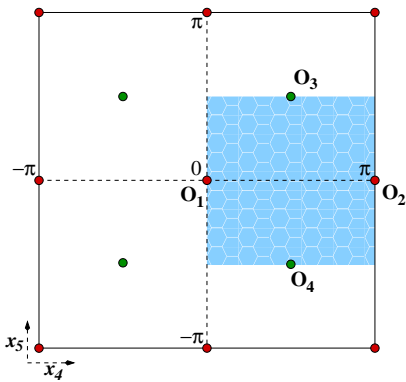}
}
\vskip1em

\caption{Fixed points of the $Z_2\times Z'_2$ orbifolding of the torus, here
represented by the whole squared in the $x_4$-$x_5$ plane. For simplicity,
the coordinates are given in units of $R$.
The shadowed region corresponds to the  actual fundamental space.}
\end{figure}

It is worth noticing that the distribution of the fixed points  is
such that the $T^2/Z_2\times Z'_2$ orbifolding breaks the
translational invariance on the torus down to a discrete
translational group, $\cal P$, which maps the equivalent fixed
points among themselves. The orbifold also keeps a  discrete $Z_4$
rotational symmetry around any fixed point. Notice also that, as it
is shown in figure 2,   the   discrete translation ${\cal P'}: \vec
y \rightarrow y'$ maps the  fundamental space into itself up to the
interchange on the $Z_2$ and $Z'_2$ fixed points:
$O_1\leftrightarrow O_3$ and $O_2\leftrightarrow O_4$, which is
equivalent to the  exchange of projections: $Z_2\leftrightarrow
Z'_2$.

\begin{figure}[ht]
\centerline{ \epsfysize=140pt \epsfbox{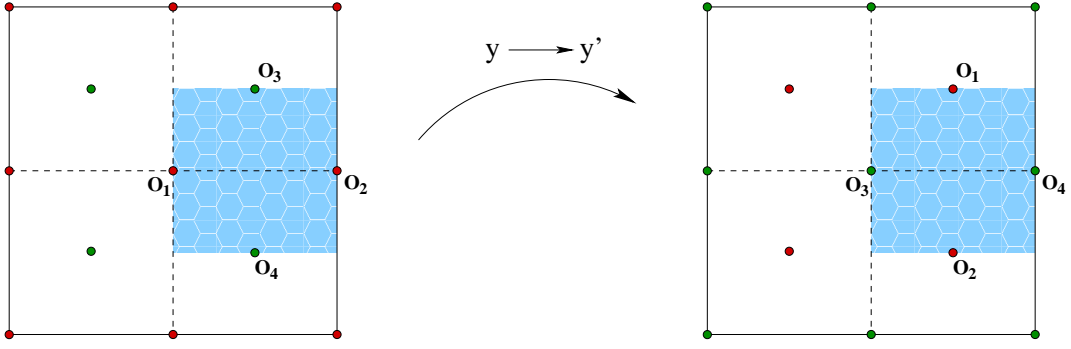} } \vskip1em

\caption{Mapping of the orbifold under the discrete translation
${\cal P'}: \vec{y}\rightarrow \vec{y}~'$. Notice that this
transformation is equivalent to a redefinition on the projections:
$Z_2\leftrightarrow Z'_2$, which interchanges the corresponding
fixed points.}
\end{figure}

The orbifold does not break completely the six dimensional Poincare
invariance down to the 4D Poincare group, but it rather keeps an
additional  discrete subgroup:
 $Z_4\times {\cal P}$, so maintaining the $Z_8$ symmetry
already existing in the $T^2/Z_2$ orbifold~\cite{ponton}.
 This is a remarkable result whose phenomenological consequences we have
already mention in section II~\footnote{All these conclusions are
true provided there are no fields attached to the fixed points,
which explicitly could break the remnant discrete Lorentz
symmetries.}: (i) the conservation of the extra momentum and (ii)
the conservation of the $U(1)_{45}$ symmetry modulus operators with
$\Delta\Sigma_{45} = 8$; which forbids Majorana neutrinos and
introduces  a large suppression for proton decay.

\subsection{Mode expansion}

Any  generic field $\varphi(x^\mu,\vec{y})$ with
given $Z_2\times Z'_2$ quantum numbers ($z,z'$), for $z,z'=\pm$,
can be  Fourier expanded as
 \be
 \varphi^{(z,z')}(x^\mu,y) =
 \sum_{n_1,n_2=0}^{\infty}
 \xi^{(z,z')}_{n_1n_2}(\vec{y})~\varphi_{n_1 n_2}(x^\mu) +
 \sum_{n_1,n_2=1}^{\infty}
 \zeta^{(z,z')}_{n_1n_2}(\vec{y})~\tilde\varphi_{n_1n_2}(x^\mu) ~.
 \label{kkexp}
 \ee
The $\xi^{(z,z')}_{n_1n_2}$ and $\zeta^{(z,z')}_{n_1n_2}$ modes,
properly normalized on the fundamental space, are given by
 \bea
 \xi^{(+,\pm)}_{n_1n_2} =
       {\eta_{\vec{n}}\over \pi R} \cos{\vec{n}\cdot \vec{y}\over R}~;
 \quad
 \zeta^{(+,\pm)}_{n_1n_2} =
       {\sqrt{2}\over \pi R}\cos{\vec{n}\cdot\sigma_3 \vec{y}\over R}~;
 \label{cos}\\[1ex]
 \xi^{(-,\pm)}_{n_1n_2} =
       {\sqrt{2}\over \pi R}\sin{\vec{n}\cdot \vec{y}\over R}~;
 \quad
 \zeta^{(-,\pm)}_{n_1n_2} =
       {\sqrt{2}\over \pi R}\sin{\vec{n}\cdot\sigma_3 \vec{y}\over R}~;
 \label{sin}
 \eea
where  both  $n_1$ and $n_2$ are either even or odd numbers for the
$(\pm,\pm)$ charged modes, whereas $n_1$ should be even (odd) when
$n_2$ is odd (even) for $(\pm,\mp)$; otherwise the modes are
identically zero. Clearly, $\vec{n}$ stands for the vector
$(n_1,n_2)$. The normalization factor $\eta_{\vec{n}}$ that appears
in  Eq.~(\ref{cos}) is equal to 1 for $\vec{n}=0$ and $\sqrt{2}$
otherwise. Notice that only fields with $(+,+)$ quantum numbers
have zero modes.

Using the expansion in Eq.~(\ref{kkexp}), it is easy now to check
that  the Lagrangian satisfies: ${\cal
L}[\varphi^{(z,z')}(x^\mu,\vec{y}~')] = {\cal
L}[\varphi^{(z',z)}(x^\mu,\vec{y})]$. That confirms our  previous
observation that the discrete translation $\vec{y}\rightarrow
\vec{y}~'$ is equivalent to a permutation on the $Z_2$ and $Z'_2$
charges. That also shows the invariance of the lagrangian under the
$Z_4$ spatial symmetry proper of the $T^2/Z_2$ orbifold.

\begin{figure}[ht]
\vskip1em \centerline{ \epsfysize=140pt \epsfbox{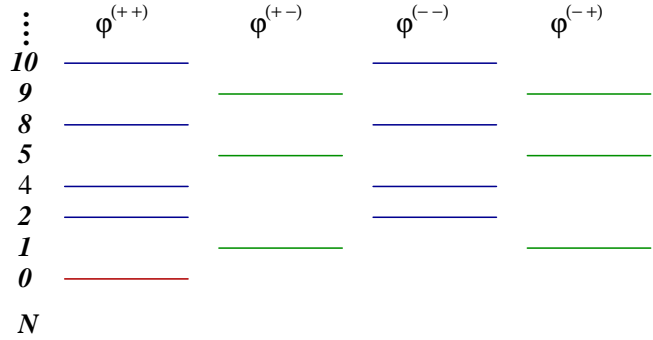} }
\vskip1em

\caption{KK mass spectrum of a generic field, $\varphi$. The KK
index $N = m_{KK}^2 R^2$. Notice that only the (++) fields develop
zero modes, and that part of the spectrum has been projected out.}
\end{figure}

In the effective 4D theory the mass of each mode has the form
 \be
 m_{N}^2 = m_0^2 + {N\over R^2}~;
 \label{mass}
 \ee
with $N=\vec{n}^2=n_1^2 + n_2^2$. In this equation $m_0$ is the
Higgs vacuum mass contribution, and the physical mass of the zero
mode. A typical spectrum of Kaluza Klein  (KK) levels is shown in
Fig. 3. Except for the zero mode, the spectrum is degenerated and
rather complex. Degeneracy for a given mass level  depends on the
array of $\{n_1,n_2\}$ numbers that give the numerical value $N$.
For arrays of the form $\{n,0\}$ (up to a permutation) and
$\{n,n\}$, the natural  degeneracy of the level is equal to 2. For
all other cases the degeneracy is equal to~4. There often are,
however, some levels with a larger accidental degeneracy due to
some numerical coincidences: some natural numbers have more than
one decomposition of the form $N= n_1^2 + n_2^2$. For instance, one
can write $25 = 5^2 + 0 = 4^2 + 3^2$. Thus, the  6-fold degeneracy
of the $N=25$ level sums over the degeneracy of both the
decompositions. Table 1. gives the degeneracy of the first KK mass
levels for the four classes of $Z_2\times Z'_2$ charges. Notice
that the spectrum is only a fourth of the one associated to $T^2$.
Indeed, the `missing' states have been projected out by the
orbifolding. It is worth noticing that the $(\pm,\pm)$ eigenmodes
lack the very first (N=1) excited mode, whereas it is present for
the $(\pm,\mp)$. Therefore the lightest KK particle on the model
will come from these last class of fields.

\begin{table}
\begin{tabular}{||c| c | c|| c | c | c ||}
  $N$  & $(\pm,\pm)$ & $(\pm,\mp)$ & $N$ & $(\pm,\pm)$ & $(\pm,\mp)$\\
\hline
1  & ---  &   2  & 13 & ---  &   4   \\
2  &  2   &  --- & 16 &  2   &  ---  \\
4  &  2   &  --- & 17 & ---  &   4   \\
5  & ---  &   4  & 18 &  2   &  ---  \\
8  &  2   &  --- & 20 &  4   &  ---  \\
9  & ---  &   2  & 25 & ---  &   6   \\
10 &  4   &  --- & 26 &  4   &  ---  \\
\end{tabular}
\caption{
Level degeneracy  according to the
$Z_2\times Z'_2$ field charges.
Here $N$ labels the physical mass level. A dash indicates a mode that has been
projected out from the spectrum by  orbifolding.
The non degenerated zero mode ($N=0$) has been
omitted in this table.}
\end{table}

\section{$Z_2\times Z'_2$ charge assignments. }
\subsection{Gauge boson charges: Orbifold symmetry breaking}

We now proceed to show the way the gauge and matter fields of our
left right model transform under the $Z_2\times Z'_2$ symmetries.
We will closely follow a similar  prescription to the one given in
Reference~\cite{5dlrm,nandi} for the case of five dimensions. We
take Gluons and $B-L$ gauge boson to transform under the $Z_2$ and
$Z'_2$  projections as follows: \bea G_\mu(x,\vec{y})\rightarrow
G_\mu(x,\vec{y})~;&\qquad &
G_{a}(x,\vec{y})\rightarrow  - G_{a}(x,\vec{y})~; \nonumber\\
B_\mu(x,\vec{y})\rightarrow B_\mu(x,\vec{y})~;&\qquad&
B_{a}(x,\vec{y})\rightarrow  - B_{a}(x,\vec{y})~;
\label{gbtrans}
\eea
for $a=4,5$.
For the $SU(2)$'s gauge bosons we define the  matrix
$W = \vec{W}\cdot \vec{\sigma}$.
The $Z_2\times Z'_2$ transformation properties of $W_{1,2}$ are then written as
\bea
W_{\mu}(x^{\mu}, \vec{y})&\rightarrow& W_{\mu}(x^{\mu}, -\vec{y})
       ~~=~\quad PW_{\mu}(x^{\mu},\vec{y}) P^{-1}~; \nonumber \\
W_{a}(x^{\mu}, \vec{y})&\rightarrow& W_{a}(x^{\mu}, -\vec{y})
       ~~=~-PW_{a}(x^{\mu},\vec{y}) P^{-1}~; \nonumber \\
W_{\mu}(x^{\mu}, \vec{y}~')&\rightarrow& W_{\mu}(x^{\mu},-\vec{y}~')
       ~=~\quad P'W_{\mu}(x^{\mu},\vec{y}~') P^{'-1}~; \nonumber \\
W_{a}(x^{\mu}, \vec{y}')&\rightarrow& W_{a}(x^{\mu}, -\vec{y}~')
       ~=~-P'W_{a}(x^{\mu},\vec{y}~')  P^{'-1}~;
\label{wtrans}
\eea
where $P$, and $P'$ are two by two diagonal matrices,
acting on the group space,  that we chose to be
(i) $P~=~P'~=~diag(1,1)$ for the $SU(2)_1$ gauge bosons; and
(ii) $P~=~diag(1,1)$ and  $P'~=~ diag(1,-1)$, for those of $SU(2)_2$.

Above transformation properties can be briefly
summarized in terms of the
following $Z_2\times Z'_2$ charge assignments:
\begin{eqnarray}
G_\mu(+,+);\quad B_\mu(+,+);\quad
W_{1,\mu}^{3,\pm}(+,+);\quad  W^3_{2,\mu}(+,+);
 \quad W^\pm_{2,\mu}(+,-); \nonumber\\
G_{a}(-,-);\quad B_a(-,-);\quad
W_{1,a}^{3,\pm}(-,-);\quad W^3_{2,a}(-,-);
 \quad W^\pm_{2,a}(-,+).
\label{gparity}
\end{eqnarray}
With these assignments, one finds that at the $Z'_2$ fixed points,
$O_{1,2}$, the charged $W^\pm_{2,\mu}$ bosons vanish. Thus, the
$SU(2)_2$ symmetry breaks down to its $U(1)_{I_{3,2}}$ subgroup,
whereas all  other group factors remain unbroken at such points. In
contrast, at the $Z_2$ fixed points all the gauge symmetry remains
intact. Nevertheless, due to the breaking of the symmetry at two of
the fixed points, in the effective four dimensional theory the
symmetry is also broken. The residual group is identified as the
one on our previous model:
 $SU(3)_c\times SU(2)_L\times U(1)_{I_{3R}}\times U(1)_{B-L}$.
Notice that this group can also be written as:
  $[SU(3)_c\times SU(2)_L\times U(1)_Y]\times U(1)_{Y'}$.
That is the SM symmetry times an extra $U(1)_{Y'}$ factor, which is
generated by the operator:
 ${1\over 2 }Y'\equiv\sqrt{\frac{2}{5}} I_{3,2}
 -\frac{3}{\sqrt{10}} {1\over2} (B-L)$.
One can easily check this statement by looking at the effective 4D
theory. One  finds that only the gauge bosons associated with this
generator have zero modes, which  are massless at this stage.

\subsection{Fermion charges}
We now turn to the fermion content. In general under the $Z_2$ and  $Z'_2$
projections a fermion transforms as
 \bea
\Psi(x^\mu,\vec{y}) &\rightarrow & \Psi(x^\mu,-\vec{y})~=~
  \epsilon~\Sigma_{45} P \Psi(x^\mu,y);  \nonumber \\
\Psi(x^\mu,\vec{y}~') &\rightarrow & \Psi(x^\mu,-\vec{y}~')~=~
  \epsilon'~\Sigma_{45} P'\Psi(x^\mu,\vec{y}~')~;
 \label{ftrans}
\eea
where $\epsilon,\epsilon' = \pm 1$ are the overall $Z_2$ and  $Z'_2$ charges
of the six dimensional fermion, respectively.
In last equation $P$ and $P'$ are the very same
matrices used in Eq.~(\ref{wtrans}) which act on  the $SU(2)_{1,2}$
group indices. They commute with the $\Sigma_{45}$ matrix that acts on the
spinorial space.
Clearly, the actual $Z_2\times Z'_2$ charges of each fermionic component
depend on the final combination of $\epsilon$'s, $P$'s and the $\Sigma_{45}$
charges.
Notice that in the chiral representation,
as it is  given in the appendix,
the $\Sigma_{45}$ operator is diagonal and has the form
$\Sigma_{45}=diag(\gamma_5,-\gamma_5)$, which indicates that
the 4D left and right handed components of a chiral 6D fermion actually
hold opposite parities under the $Z_2$ and  $Z'_2$ projections.
In fact, using the same chiral representation of the appendix,
Eq.~(\ref{ftrans}) can be explicitly decomposed in terms of  left and
right handed components,  to  get the simplified transformation rules
$\Psi_{\pm,R} \rightarrow \pm \epsilon P \Psi_{\pm,R}$ and
$\Psi_{\pm,L} \rightarrow \mp \epsilon P \Psi_{\pm,L}$ for $Z_2$, and a
similar expression for $Z'_2$.

We take the following ($\epsilon,\epsilon'$)
assignments for the matter content of the model:
${\cal Q}_1(+,+)$;  ${\cal Q}'_1(+,-)$;
${\cal \psi}_1(+,+)$; ${\cal \psi}'_1(-,+)$;
${\cal Q}_2(+,+)$;  ${\cal Q}'_2(+,-)$;
${\cal \psi}_2(+,-)$; and  ${\cal \psi}'_2(-,-)$.
Combining this choice with the one made for the $P$ and $P'$ matrices
in Eq.~(\ref{wtrans}), it is easy to see that the various fermion
representations get the following $Z_2\times Z'_2$ charges, for quarks:
\bea
 Q_{1,L}\equiv
   \left(\begin{array}{c} u_{1L}(+,+)\\ d_{1L}(+,+)\end{array}\right);
 &\quad&
 Q'_{1,L}\equiv
   \left(\begin{array}{c} u'_{1L}(+,-)\\ d'_{1L}(+,-)\end{array}\right);
   \nonumber \\
 Q_{1,R}\equiv
   \left(\begin{array}{c} u_{1R}(-,-)\\ d_{1R}(-,-)\end{array}\right);
 &\quad&
 Q'_{1,R}\equiv
   \left(\begin{array}{c} u'_{1R}(-,+)\\ d'_{1R}(-,+)\end{array}\right);
   \nonumber \\ [1ex]
 Q_{2,L}\equiv
   \left(\begin{array}{c} u_{2L}(-,-)\\ d_{2L}(-,+)\end{array}\right);
 &\quad&
 Q'_{2,L}\equiv
   \left(\begin{array}{c} u'_{2L}(-,+)\\ d'_{2L}(-,-)\end{array}\right);
   \nonumber \\
 Q_{2,R}\equiv
   \left(\begin{array}{c} u_{2R}(+,+)\\ d_{2R}(+,-)\end{array}\right);
& \quad&
Q'_{2,R}\equiv
   \left(\begin{array}{c} u'_{2R}(+,-)\\ d'_{2R}(+,+)\end{array}\right);
\label{quarks}
\eea
and for leptons:
 \bea
 \psi_{1,L}\equiv
   \left(\begin{array}{c} \nu_{1L}(+,+)  \\ e_{1L}(+,+)\end{array}\right);
  &\qquad&
 \psi'_{1,L}\equiv
   \left(\begin{array}{c} \nu'_{1L}(-,+)  \\ e'_{1L}(-,+)\end{array}\right);
   \nonumber \\
 \psi_{1,R}\equiv
    \left(\begin{array}{c} \nu_{1R}(-,-)  \\ e_{1R}(-,-)\end{array}\right);
  & \qquad &
 \psi'_{1,R}\equiv
  \left(\begin{array}{c} \nu'_{1R}(+,-) \\ e'_{1R}(+,-)\end{array}\right);
   \qquad
   \nonumber \\ [1ex]
 \psi_{2,L}\equiv
   \left(\begin{array}{c} \nu_{2L}(-,+) \\ e_{2L}(-,-)\end{array}\right);
  &\qquad&
 \psi'_{2,L}\equiv
    \left(\begin{array}{c} \nu'_{2L}(+,+)\\ e'_{2L}(+,-)\end{array}\right);
  \nonumber \\
 \psi_{2,R}\equiv
    \left(\begin{array}{c} \nu_{2R}(+,-)\\ e_{2R}(+,+)\end{array}\right);
&   \qquad &
 \psi'_{2,R}\equiv
  \left(\begin{array}{c} \nu'_{2R}(-,-)\\ e'_{2R}(-,+)\end{array}\right).
\label{leptons}
 \eea
The zero mode fermion content of the model is the same as the
standard model plus an additional sterile neutrino per family. Note
that  the zero mode spectrum is actually the one of the model
discussed in section II. From now one can identify  the fermion
fields having zero modes with the self-explaining standard
notation: $Q, L, u_R, d_R, e_R$ and $\nu_s$.

\subsection{Scalar content: Spontaneous symmetry breaking.}

The effective 4D gauge symmetry $SU(2)_L\times U(1)_Y\times U(1)_{Y'}$
of the model is being
spontaneously broken by an appropriate set of Higgs fields.
The minimal set required for this purpose as well to give masses to the
SM fermions was introduced in Ref.~\cite{5dlrm}.
It has a bidoublet $\phi(2,2,0)$ and doublets $\chi_L(2,1,-1)$ and
$\chi_R(1,2,-1)$ with the following $Z_2\times Z'_2$ quantum numbers:
\be
\phi \equiv
\left(\begin{array}{cc} \phi^0_u(+,+) & \phi^+_d(+,-)\\
   \phi^-_u(+,+) &  \phi^0_d(+,-)\end{array}\right);\quad
\chi_L\equiv \left(\begin{array}{c} \chi^0_L(-,+) \\
   \chi^-_L(-,+)\end{array}\right); \quad
\chi_R\equiv \left(\begin{array}{c} \chi^0_R(+,+) \\
   \chi^-_R(+,-)\end{array}\right) .
 \ee
At the zero mode level, only the SM doublet $(\phi^0_u, \phi^-_u)$ and a
singlet $\chi^0_R$ appear. The vacuum expectation values (vev's)
of these fields,
namely  $\langle\phi^0_u\rangle = v_{wk}$ and
$\langle\chi^0_R\rangle= v_R$, break the SM symmetry and the extra $U(1)_Y'$
gauge group, respectively. As a matter of fact,  in the six dimensional theory,
$\langle\chi^0_R\rangle$ breaks the group  $SU(2)_2\times U(1)_{B-L}$
down to $U(1)_Y$, given a universal mass contribution to all KK $W_R$
modes.

\section{Mass spectrum and phenomenology.}

The analysis of the masses spectrum follows almost identically the
one already presented in Ref.~\cite{5dlrm}, with the KK masses now
given accordingly to Eq.~(\ref{mass}). Vacuum contribution to the
masses of all particles  is independent of the KK number and
directly calculable in the six dimensional theory. Therefore, it
introduces a global shifting of the KK spectrum by fixing the value
of $m_0$, in Eq.~(\ref{mass}),  for every each field. Mixings in
the theory are only produced trough the vacuum, and they are also
six dimensional. Thus, no mixing among fields with different KK
number is possible. Briefly, these are our main results:

There is no mixing among charged gauge bosons ($W^\pm_{L,R}$). The
$W_L$ zero mode  mass is, as usual, $m_{0,W_L}^2=M_{W_L}^2 = g_L^2
v_{wk}^2/2$ whereas,  $W_R$ gets $m_{0,W_R}^2 = {g_R^2 \over
2}\left( v_R^2 + v_{wk}^2 \right)$. Here, $g_{L,R}$ represent the
gauge coupling constants of $SU(2)_{L,R}$ respectively. Notice,
that due to its $Z_2\times Z'_2$ charges, $W_R$ has no zero mode,
and thus, its  lower mode gets a $1/R^2$ KK mass contribution.
Without mixings in this sector, added to KK conservation, most of
the former 4D constraints on  $v_R$ disappear. For instance: there
are no new  tree level contributions to muon decay. Moreover, $W_R$
has not tree level contribution to double beta decay, nor a
relevant contribution  on the $K-\bar K$ mixing (last comes out
very suppressed).

On the neutral sector photon decouples
and  remain massless.
the other two neutral gauge bosons mix among themselves.
For $v_R\gg v_{wk}$,
the mass of the standard $Z$ boson gets a mass correction due to this mixing:
$M_Z^2 = m_Z^2 - \delta m_Z^2$,
which in the symmetric limit ($g_L=g_R$) has the form
\be
{\delta m_Z^2\over m_Z^2}\approx
{\cos^2 2\theta\over \cos^4\theta}\left({v_{wk}\over v_R}\right)^2~,
\ee
whereas the mixing is given as
\be
\tan\beta\approx
 \left({v_{wk}\over v_R}\right)^2
{(\cos 2\theta)^{3/2}\over \cos^4\theta}~. \ee In these equations
$\theta$ corresponds to the standard weak mixing angle. Existence
of a mixing in the neutral currents imposes a lower limit on $v_R$
that can be calculated as in four dimensions due to the KK
conservation. On gets $v_R\gs 800$~GeV. Limits on $R$ are very
weak. Production of KK pair excitations of the $W_L$ in
colliders~\cite{cheng} imposes a limit that ranges from 400 to 800
GeV. The first KK level for all these neutral fields get a  KK mass
contribution equal to $2/R^2$. They lack the very first possible
level of the tower.

The most general Yukawa couplings in the model are
\be
h_u \bar Q_1\phi Q_2 + h_d \bar Q_1\tilde\phi Q'_2 +
h_e\bar \psi_1\tilde \phi \psi_2 +
h'_u \bar Q'_1\phi Q'_2 + h'_d \bar Q'_1\tilde\phi Q_2 +
h'_e\bar \psi'_1\tilde \phi \psi'_2 +
 h.c.;
\label{Yuk} \ee where $\tilde \phi\equiv \tau_2\phi^* \tau_2$ is
the charge conjugate field of $\phi$. A six dimensional realization
of the left-right symmetry, which interchanges  the subscripts:
$1\leftrightarrow 2$, is obtained provided the $3\times 3$ Yukawa
coupling matrices satisfy the constraints: $h_u~=~h^{\dagger}_u$;
$h'_u~=~h^{'\dagger}_u$; $h_e~=~h^{\dagger}_e;$
$h'_e~=~h^{'\dagger}_e;$ $h_d~=~h^{'\dagger}_d$. At the zero mode
level one obtains the SM Yukawa couplings \be {\cal L}~=~h_u \bar
Q\phi_u u_R + h_d \bar Q\tilde\phi_u d_R + h_e \bar L\tilde \phi_u
e_R + h.c. \label{yukawa} \ee It is important to notice that in the
above equation $h_{u,e}$ are hermitian matrices, while $h_d$ is
not. Therefore, whereas $h_{u,e}$ are diagonalizable by a single
unitary matrix, $h^{diag}_{u, e} = V_{u,e} h_{u,e}
V^{\dagger}_{u,e}$; $h_d$ needs two of such matrices: $h^{diag}_{d}
~=~ V_d h_d U^{\dagger}_d$. Last implies that, unlike the case of
standard left-right models, the left and right handed quark mixings
are different from each other. Indeed, for left handed quarks one
gets the CKM matrix $U_{CKM}~=~V^{\dagger}_uV_d$; while the
corresponding right handed charged current mixing matrix for quarks
is $U^{R}~=~V^{\dagger}_uU_d$.

Considering the decomposition $\bar\Psi_1\Psi_2=
\bar\Psi_{1,L}\Psi_{2,R} + \bar\Psi_{1,R}\Psi_{2,L} $, one can
easily read the fermion masses induced by vacuum. Particularly, we
find that left and right handed chiral partners (in same
representation) have the same mass out of the Yukawa couplings.
Also, we stress that all but the neutrino fields, $\nu_1$;
$\nu'_1$, $\nu_2$; and $\nu'_{2}$, get a mass contribution
proportional to $v_{wk}$. Standard and sterile neutrinos remain
massless at this point. They will get small masses, however,
through non renormalizable operators as we will discuss next. It is
worth noticing that being  neutrinos the particles with the smaller
vacuum mass correction, the lightest KK fermionic particle in the
model will be a  KK neutrino: the ones associated to $\nu'_1$ and
$\nu_2$, which have the lowest possible KK mass level with a mass
$1/R$.

\section{Neutrino masses}

As already mentioned in section V, the orbifolding leaves a
residual discrete $Z_8$ Lorentz symmetry acting on spinors, which
is generated  by $\Sigma_{45}$. The conservation of such a symmetry
constrains the possible bilinear couplings among fermions.
Particularly, as already mentioned, the symmetry forbids a Majorana
self coupling for chiral fields. A simple way of understanding this
is by noticing that  the charge conjugation does not change the 6D
chirality  (see the appendix), thus, $\bar\Psi^c_\pm~\Psi_\pm$ has
$\Delta \Sigma_{45} =2$. This is troublesome for understanding the
smallness of the neutrino mass since without a large Majorana mass,
the see-saw mechanism is not any more at work. The problem was
already noticed for the six dimensional SM, and an alternative
solution  was explored in Ref.~\cite{ponton2}. Such a solution,
however, makes use of singlet bulk neutrinos propagating in a
larger (7D) space where the additional dimension is
warped~\cite{rs}. This was similar in spirit to the  model of
Ref.~\cite{grossman}.\footnote{ For a different approach to
neutrino mass problem in 5D left-right models, see~\cite{perez}.}

 As we already remarked in the last section, in our model
a Dirac neutrino mass does not come at the renormalizable level of
the theory. The model, however, contains neutrinos in both the
chiral sectors, and thus a mass term can be written in the
effective theory through non renormalizable operators. Such an
operator, of course, should be $Z_8$ invariant, so it  ought to
have $\Delta \Sigma_{45} =0$. Looking at  the zero mode spectrum of
the model we find that all the SM fields ($Q, L, u_R, d_R, e_R$)
have $\Sigma_{45}$ charge equal to $1$, whereas $\nu_s$ has
$\Sigma_{45}=-1$. Therefore the only possible neutrino mass term at
the zero mode level is of the form $\nu^T_{sL}C^{-1} \nu_L$. As in
the previous example, this  implies that the neutrino is a Dirac
particle. The lowest order operator that contains such a term is of
mass dimension eleven:
 \be
 {h\over M_*^{5}} \bar\psi_1\tilde{\phi}{~\psi'_2}^c
\chi_R\chi_R~,
 \ee
which  is equivalent to the operator considered in
Eq.~(\ref{nuop}). It generates at the 4D effective theory the
suppressed dimension six operator~\cite{pires}
 \[{h\over  (M_*\pi R)^3}{{\bar L \phi_u \nu_s^c}\chi_R^2\over M_*^2}~. \]
which, for $v_R\approx 1/\pi R\approx 1$~TeV; $M_*\approx 100$~TeV,
and a coupling strength of order 0.1
gives $m_\nu\sim 1$~eV.

It is remarkable that the model does not need the introduction of
singlet bulk neutrinos. Nevertheless, at least one warped or several flat
extra dimensions  (not seen by the model fields)
may be used (though not mandatory) to compensate for the gap between the
compactification $R$ and fundamental scale, $M_*$. The value of $M_*$
we are considering  is consistent with current bounds on graviton effects (see
for instance Refs.~\cite{exp1}).

\section{Baryon non conservation}
Turning to the question of proton decay, the considerations are very
similar to the $SU(2)_L\times U(1)_{I_{3R}}\times U(1)_{B-L}$ model
discussed in section II. The residual discrete $Z_8$
spatial symmetry of the orbifold naturally constrains the  proton life time.
All allowed  baryon number violating now have to be invariant under the
full left-right group and are suppressed down to levels consistent with
the experiment as in the $SU(2)_L\times U(1)_{I_{3R}}\times U(1)_{B-L}$
model of section II.

The lowest dimension $\Delta B=1$ operators,
invariant under gauge and orbifold symmetries in this case are:
{\it  (i) dimension 16 operators:}
 \bea
a)&\qquad&(\bar\psi_1  Q_2)
 (\bar Q_1^c \psi'_2)(\bar Q_2^c \nt{D} \psi'_2)~; \nonumber \\
b) &&   (\bar\psi_1 \nt{D} Q_1)
(\bar Q_2^c \Gamma^M \psi'_2)(\bar Q_2^c \Gamma_M \psi'_2)~; \nonumber \\
c)&& (\bar\psi_2 \nt{D} Q_2)
 (\bar Q_2^c \Gamma^M \psi'_2)(\bar Q_2^c \Gamma_M \psi'_2)~;
 \label{dim16}
 \eea
where we have considered operators with the smallest number of
$\Gamma$'s. The insertion of any product of $\Sigma_{MN}$ matrices
will not introduce any new operator to the effective 4D theory.
Permutation of $\nt{D}$ and $\Gamma_M$ is also allowed. The overall
mass suppression on these couplings is of order $M^{-10}$. Thus, at
the 4D effective theory one gets the six fermion couplings:
 \bea
a) &\qquad & (\bar L d_R)(\bar Q^c \nu_s) (\bar d_R^c \nt D \nu_s)~; \nonumber \\
b) && (\bar L\nt D Q)
     (\bar d_R^c \gamma^\mu \nu_s)(\bar d_R^c \gamma_\mu \nu_s)~; \nonumber \\
c) && (\bar e_R \nt D d_R)
   (\bar d_R^c \gamma^\mu \nu_s)(\bar d_R^c \gamma_\mu \nu_s)~;
 \eea
respectively, with and overall suppression $\sim [M_*^6 (M\pi R)^4]^{-1}$.
The decay modes and predictions for nucleon lifetimes are same as in
section II and we do not repeat the discussion here.

\section{Two component Kaluza-Klein dark matter}

In this section, we look at possible dark matter candidates in our
model. The fact that in universal extra dimension models with a low
compactification scale, KK excitations of photon is stable and can
act as a dark matter candidate has been discussed
recently\cite{feng}. This is however dependent on the nature of the
orbifold compactification and in models with $S^1/Z_2$ orbifold,
$\gamma^1_{KK}$ is clearly the only possibility. The main reason
for this is that, of all the particles at the first KK level, the
one with the slowest annihilation rate is the first KK excitation
of the photon. First KK excitations of particles such as neutrino
or electron which have antiparticles, can annihilate via the
exchange of zero mode states (n=0 states) (such as the usual
Z-boson) whereas the annihilation of the KK modes of photon can
proceed only via the exchange of another KK excitation due to
conservation of ``fifth'' momentum. The latter annihilation cross
section is therefore highly suppressed compared to the annihilation
rate other particles in the theory. This in turn implies that the
$\gamma^{1}_{KK}$ will be present in the late universe in greater
abundance than all other KK modes and can play the role of dark
matter of the Universe.

When the compactification orbifold is different such as
$T^2/(Z_2\times Z'_2)$ as it is  in the left-right symmetric case
of this paper, the situation changes drastically. In this case
there are two classes of levels: one class corresponding to even KK
number with $Z_2\times Z'_2$ quantum numbers $(+,+)$ and $(-,-)$
and another class corresponding to odd KK number, corresponding to
$Z_2\times Z'_2$ quantum numbers $(+,-)$ and $(-,+)$. Of these only
$(+,+)$ modes contain the zero mode (see Fig. 3). This implies that
the lightest KK modes are those in $(+,-)$ or $(-,+)$ class. In
these theories the mass of the lowest $\gamma_{KK}$ mode is twice
the mass of the lowest KK modes of states $(\pm,\mp)$ type.
Therefore the discussion of dark matter candidate has to take this
into account to see which particle really is the dark matter. What
we find is that the role of dark matter is shared by two different
particles: $\gamma^1_{KK}$ and $\nu^1_{2,KK}$ because the
annihilation rate of both of these states are mediated by particles
with masses of order $R^{-1}$ and are therefore comparable. In the
case of $\gamma^1_{KK}$, the exchange particles are the lowest KK
modes of the standard model particles and in case of
$\nu^1_{2,KK}$, they are the $Z'$ gauge boson. Below we give
semi-quantitative arguments to show that their abundances in the
present universe are comparable to each other. The dark matter of
the universe should therefore have two components.

To get $\Omega_X$ where $X$ denotes either of the above particles, we need
their present number density $n_0(X)$ and mass. To calculate
$n_0(X)$, note that if the X-particle decouples from the
Hubble
expansion at a temperature $T^*$, after decoupling, the ratio
$n(X,T^*)/n(\gamma, T^*)$ does not change except for a predictable
fraction  which correspond to contributions from particle annihilation
to photon density. If we call that fraction $\beta$, the present number
density of $n_0(X)$ can be written as:
\begin{eqnarray}
n_0(X)\simeq \frac{n(X,T^*)}{n(\gamma,T^*)}\beta n_0(\gamma)
\end{eqnarray}
For a given particle $X$, we can roughly estimate
$\frac{n(X,T^*)}{n(\gamma,T^*)}$ as follows:
\begin{eqnarray}
\frac{n(X,T^*)}{n(\gamma,T^*)}~=~\frac{1.2 \sqrt{g^*}}{<\sigma_X v>}
\frac{(T^*)^2}{M_{P\ell}}
\end{eqnarray}
where $\sigma_X$ denotes the annihilation rate for the pairs of $X$
particles. Thus the relative abundance of the different particles is
determined by the magnitude of $\sigma_X$, which we estimate below.

For $\gamma^1_{KK}$, $\sigma_X$ is given by
\begin{eqnarray}
\sigma_{\gamma^1_{KK}}\simeq \frac{e^4 n_f}{16\pi M^2}
\end{eqnarray}
where $M=R^{-1}$ and $n_f=\sum_i Q^4_i$, where $Q_i$ is the electric
charge of the final state particle; $n_f\simeq 6$ for our model, where
states contributing are all charged leptons (zero modes), quarks and
$W^{\pm}$.

On the other hand for $\nu^1_{2, L, R}$, the annihilation cross sections
are given by:
\begin{eqnarray}
\sigma_{\nu_2}\simeq \left(\frac{g_2^2}{16 cos^2\theta_W
M^2_{Z'}}\right)^2
2n'_f
\end{eqnarray}
where $g_2$ is the weak $SU(2)$-gauge coupling constant. Putting in
the values for $g^2_2\simeq 0.41$ and $sin^2\theta_W\simeq 0.23$
and taking $M_{Z'}$ close to $R^{-1}$, we estimate that
$\sigma_{\nu_2}\simeq \sigma_{\gamma_{KK}}$. There are two $\nu_2$
states (L and R); however, the $\gamma_{KK}$ is twice as massive as
the $\nu_2$. therefore all three states have nearly same
contribution and should equally share the dark matter role. Again
one can give a rough evaluation of the contribution of each
particle to $\Omega_X$ and we find
 \begin{eqnarray}
 \Omega_X\simeq \left(\frac{M}{200~GeV}\right)^2 5~ KeV
 \end{eqnarray}
This contribution is roughly of the same order of magnitude as required to
give 50\% of the critical mass density. Thus we see that the dark matter
of the universe in our model has two components to it. This
feature should
have implications for dark matter detection as well as dark matter
annihilation in galaxies. A detailed analysis of these questions will be
the subject of a forthcoming investigation.

\section{Conclusions}

In summary, we have considered local $B-L$ extensions of TeV scale
gravity models in six space-time dimensions. We have shown that two
problems of TeV scale gravity models having to do with proton decay
and neutrino mass can be solved simultaneously with two extra space
dimensions compactified on a $T^2/Z_2$ or $T^2/(Z_2\times Z'_2)$
orbifold. We have shown how this occurs in two examples: one with
the gauge group $SU(2)_L\times U(1)_{I_{3R}}\times U(1)_{B-L}$ and
another with its left-right symmetric extension. We show that the
model can have a compactification scale of order of a TeV and
fundamental scale of order 30 to 100 TeV without contradiction with
any known phenomenology. The model predicts several interesting
decay modes of both the neutron and the proton which under certain
circumstances can be within the accessible range of planned proton
decay search experiments. We also show that the dark matter of the
universe in this picture consists of two components.

\acknowledgements

The work of R. N. M. is supported by the National Science
Foundation Grant No. PHY-0099544.
 A.P.L. would like to thank the members of the High Energy Section of
the ICTP for their warm hospitality during the last two years.

 \break

\appendix
\section{}
In a six dimensional space, the $\Gamma^M$
Dirac matrices are eight by eight matrices, which
can be built from the well known $\gamma^\mu$ of the four dimensional
representation. Consider the following  possible realization:
\beas
\Gamma^\mu = \gamma^\mu\otimes \sigma_1 =
            \left(\ba{cc}
            0 & \gamma^\mu \\
            \gamma^\mu  & 0\\
              \ea\right)~;
\qquad
\Gamma^4 = i\gamma_5\otimes \sigma_1 =
           \left(\ba{cc}
            0 & i\gamma_5 \\
            i \gamma_5  & 0\\
               \ea\right)~;
\eeas
\be
\Gamma^5 = 1_{4\times 4}\otimes i\sigma_2 =
           \left(\ba{cc}
            0 & 1_{4\times 4} \\
            -1_{4\times 4}  & 0\\
               \ea\right)~;
\ee
where $\gamma^5 = -i\gamma^0 \gamma^1 \gamma^2 \gamma^3$.
It is then easy to see that the above matrices satisfy  Clifford
algebra: $\{\Gamma^M,\Gamma^N\} = 2 g^{MN}$ for $M,N=0,1,\dots,5$ and
$g^{MN} = diag(1,-1,-1,-1,-1,-1)$, as required.

In this representation one gets
\be
\Gamma^7 = \Gamma^0 \Gamma^1\cdots \Gamma^5 =
          \left(\ba{cc}
            1_{4\times 4} & 0 \\
            0 & -1_{4\times 4}\\
               \ea\right)~;
\ee
whereas the generator of rotations in the $x_4-x_5$ plane, that is the
$U(1)_{45}$ Lorentz subgroup,  is given by
\be
\Sigma_{45} = {i\over 2}~ [\Gamma^4,\Gamma^5 ] ~= ~
 \left(\ba{cc}
            \gamma_5 & 0 \\
           0 & -\gamma_5\\
               \ea\right)~.
\ee
Notice that $[\Sigma_{45},\Gamma^7] =0$; thus, one can further classify the
fermion components in terms of the $\Sigma_{45}$ eigenstates, that is
through out the $U(1)_{45}$ charge.

Furthermore, we take the four dimensional Dirac matrices in its well known
chiral representation:
 \be
 \gamma^\mu = \left(\ba{cc}
            0 & \sigma^\mu \\
            \bar\sigma^\mu & 0\\
               \ea\right)~;
 \ee
for $\sigma^\mu= (1_{2\times 2},\vec\sigma)$, with $\vec\sigma$ the Pauli
matrices, and $\bar\sigma^\mu= (1_{2\times 2},-\vec\sigma)$.
In this representation $\gamma_5$ has a totally  diagonal form:
$\gamma_5= diag(1_{2\times 2},-1_{2\times 2})$.

$\Gamma^7$ has two chiral eigenstates defined as
$\Psi_+$ and $\Psi_-$.
Clearly, any  six dimensional fermion, $\Psi$, is decomposed in the chiral
representation in terms of its
chiral components as
\[
\Psi = \left(\ba{c} \Psi_+\\ \Psi_- \ea\right)~.
\]
Let us stress that
six dimensional chirality does not correspond to four dimensional chirality.
Six dimensional chiral fields, indeed, correspond to four component fermions,
that still contain two 4D chiral components, identified as usual
as left (L) and right (R) handed components, eigenstates of $\gamma^5$:
\[
\Psi_\pm = \left(\ba{c} \psi_{\pm,R}\\ \psi_{\pm,L} \ea\right)~.
\]

Charge conjugated fields are defined as usual by the transformation
$\Psi^c = C \bar \Psi^T$, where $C$ is such that $(\Gamma^M)^T = -
C \Gamma^MC^{-1}$. It is easy to check  that in the chiral 6D
representation $C= \Gamma^0\Gamma^2\Gamma^4$. The operator $C$ also
satisfies  the identities $C^\dagger = C^{-1}= -C$. Notice also
that  because $\{C,\Sigma_{45}\} =0$, charge conjugation  changes
the  $U(1)_{45}$ charge. Moreover, as  $[C,\Gamma^7]=0$, thus
charge conjugation does not affect the six dimensional chirality.
Specifically we have $(\Psi_\pm)^c =(\Psi^c)_\pm$, in contrast to
what happens in four dimensions. The straightforward implication of
this property is the absence of a Majorana mass term for a single
chiral fermion. Indeed, the corresponding  Lorentz invariant
bilinear  that one can write with a single field and its conjugate
needs both its chiral components to couple in the form:
 \[
 \bar\Psi^c~\Psi =
 \bar\Psi^c_+~\Psi_- + \bar\Psi_-^c~\Psi_+~.
 \]
Actually this is just as in the case of the Dirac mass couplings, which,
as in four dimensions, also need both the chiralities to be written:
 \[
 \bar\Psi~\Psi =
 \bar\Psi_+~\Psi_- + \bar\Psi_-~\Psi_+~.
 \]


\end{document}